\documentstyle[preprint,aps]{revtex}
\begin{document}
\draft
\title{Enhancement of parametric pumping due to Andreev reflection}

\author{Jian Wang $^1$, Yadong Wei$^1$, Baigeng Wang$^1$, Hong Guo$^2$}

\address{1. Department of Physics, The University of Hong Kong, 
Pokfulam Road, Hong Kong, China\\
2. Center for the Physics of Materials and Department 
of Physics, McGill University, Montreal, PQ, Canada H3A 2T8}
\maketitle

\begin{abstract}
We report properties of parametric electron pumping in the presence of a 
superconducting lead. Due to a constructive interference between the 
direct reflection and the multiple Andreev reflection, the pumped current 
is greatly enhanced. For both quantum point contacts and double barrier 
structures at resonance, we obtain exact solutions in the weak pumping 
regime showing that $I_p^{NS} = 4 I_p^N$, which should be compared with 
the result of conductance $G_{NS} = 2G_N$. Numerical results are also 
provided for the strong pumping regime showing interesting Andreev assisted
pumping behaviour.
\end{abstract}


Current can flow under zero bias when two system parameters of a 
nanostructure are varied in a cyclic fashion. The physics of 
this parametric electron pump has been analyzed by 
several authors\cite{brouwer,zhou}. Recently, it has been
realized experimentally by switkes {\it et. al.}\cite{switkes} confirming
many of the theoretical predictions. So far, investigations on
parametric pumping are focused on normal nanostructures. It will be 
interesting to study a hybrid structure where a superconducting lead is 
present. In the presence of normal conductor-superconductor (NS) 
interface, an incoming electron-like excitation can be Andreev reflected 
as a hole-like excitation. This current doubling effect gives rise to the
relation for conductance $G_{NS}=2G_N$ for hybrid quantum point contacts
and quantum dots at resonance\cite{beenakker}. For dirty NS contacts, 
de Jong and Beenakker\cite{jong} found that the shot noise at subgap 
voltages is also doubled with respect to its value in normal state which 
has been confirmed experimentally\cite{experiment}. To further explore how 
does Andreev reflection modify quantum interference in the normal 
state, we have investigated the parametric pumping phenomenon in the 
presence of a superconducting lead. We find that due to quantum 
interference of the direct reflection and the multiple Andreev 
reflection, the pumped current is greatly enhanced. For quantum point 
contact and quantum dot at resonance, we obtained a relation for the 
pumped current, $I_p^{NS} = 4I_p^N$ in the weak pumping regime. 
Numerical results are presented in the strong pumping regime showing 
interesting Andreev assisted pumping behaviour which can be verified 
experimentally.

We consider a parametric pump which consists of a double barrier 
tunneling structure attached to a normal left lead and a
superconducting right lead. The double barrier structure is modeled by 
potential $V(x)=V_1\delta(x+a) +V_2\delta(x-a)$ where $V_1=V_0+V_p 
\sin(\omega t)$ and $V_2 = V_0 + V_p \sin(\omega t +\phi)$ and $V_p$ is 
the pumping amplitude. We further apply a gate voltage $v_g$ to control 
the energy level of the structure. The units are fixed by setting 
$\hbar=2m=q=1$ in the following analysis\cite{foot4}. At low frequencies, 
the adiabatic pumped current in the presence of superconducting
leads is\cite{brouwer,foot1}
\begin{equation}
I_p^{NS}=\frac{\omega}{\pi} \int_0^\tau d\tau [\frac{dN_L}{dV_1} 
\frac{dV_1}{dt} + \frac{dN_L}{dV_2} \frac{dV_2}{dt}] 
\label{current1}
\end{equation} 
where the quantity $dN_L/dV$ is the injectivity\cite{buttiker,jwang}
given, at zero temperature, by
\begin{eqnarray}
\frac{dN_L}{dV_j}= \frac{1}{2\pi} Im [S^*_{ee} \partial_{V_j} S_{ee}
- S^*_{he} \partial_{V_j} S_{he}] 
\label{inj}
\end{eqnarray}
where the first term is the injectivity of electron due to the
external parameter\cite{buttiker,jwang}, {\it i.e.} the partial density 
of states (DOS) for an electron coming from left lead and exiting the 
system as an electron,
and the second term is the injectivity of a hole, 
{\it i.e.} the DOS for a hole coming from left lead and exiting the
system as an electron.

For the hybrid nanostructure, the scattering matrix $S_{ee}$ and $S_{he}$ 
are given by\cite{beenakker,lesovik}
\begin{equation}
\hat{S} = \hat{S}_{11} + \hat{S}_{12} (1 - \hat{R}_I
\hat{S}_{22})^{-1} \hat{R}_I \hat{S}_{21}
\label{lesovik}
\end{equation}
where $\hat{S}$ is a $2\times 2$ scattering matrix for NS structure
with matrix element $S_{\mu \nu}$ ($\mu, \nu = e, h$) 
and $\hat{S}_{ij}(E)$ ($i,j=1,2$) is a diagonal $2 \times 2$ scattering 
matrix for double barrier structure with matrix element $S_{ij}(E)$ and
$S^*_{ij}(-E)$. $\hat{R}_I=\alpha \sigma_x$ is the $2\times 2$ 
scattering matrix at NS interface with off diagonal matrix element 
$\alpha$. Here $\alpha = (E-i\nu \sqrt{\Delta^2-E^2})/\Delta$ with 
$\nu=1$ when $E>-\Delta$ and $\nu=-1$ when $E<-\Delta$. In
Eq.(\ref{lesovik}), the energy $E$ is measured relative to the chemical
potential $\mu$ of the superconducting lead.  Eq.(\ref{lesovik}) 
has clear physical meaning\cite{lesovik}. The first term is the
direct reflection from the normal scattering structure and the second
term can be expanded as $\hat{S}_{12} \hat{R}_I \hat{S}_{21} + \hat{S}_{12} 
\hat{R}_I \hat{S}_{22} \hat{R}_I \hat{S}_{21} + ...$ which is
clearly the multiple Andreev reflection in the hybrid structure.
From Eq.(\ref{lesovik}) we obtain the well known expressions for the
scattering matrix $S_{ee}$ and $S_{he}$\cite{beenakker}
\begin{equation}
S_{ee}(E) = S_{11}(E) + \alpha^2 S_{12}(E) S_{22}^*(-E) M_e S_{21}(E)
\label{see}
\end{equation}
and
\begin{equation}
S_{he}(E) = \alpha S_{12}^*(-E) M_e S_{21}(E)
\label{she}
\end{equation}
with $M_e = [1- \alpha^2 S_{22}(E) S_{22}^*(-E)]^{-1}$.
In the following, we first present the exact result for
the pumped current in the weak pumping regime and then study the 
general situation numerically. In the weak pumping 
regime, Eq.(\ref{current1}) can be expanded to the lowest order in $V_p$,
\begin{equation}
I_p^{NS} = \frac{\omega \sin\phi V_p^2}{\pi} Im 
[\partial_{V_1} S_{ee}^* \partial_{V_2} S_{ee} - 
\partial_{V_1} S_{he}^* \partial_{V_2} S_{he}]
\label{i1}
\end{equation}
as compared with the expression for the normal 
structure\cite{brouwer,foot2},
\begin{equation}
I_p^{N} = \frac{\omega \sin\phi V_p^2}{\pi} Im 
[\partial_{V_1} S_{11}^* \partial_{V_2} S_{11} + 
\partial_{V_1} S_{21}^* \partial_{V_2} S_{21}]
\label{i2}
\end{equation}
where we set $V_p=0$ in $S_{\nu \mu}$ and $S_{ij}$ after the partial 
derivatives. We further assume that the Fermi energy is close to the 
chemical potential of superconducting lead, so $E \sim 0$ and $\alpha 
\sim -i$. Under this condition, $S_{he}$ and hence $\partial_{V_j} S_{he}$ 
are pure imaginary numbers for general $V_j(t)$. As a result, $S_{he}$ 
does not contribute to $I_p^{NS}$ in Eq.(\ref{current1}) as long as 
$E=0$\cite{foot3}. The only contribution comes from $S_{ee}$ which is the 
superposition of the direct reflection and the multiple Andreev reflection.
We will consider two cases: (a) a quantum point contact, {\it e.g.} 
$V_0=0$; and (b) the double barrier quantum dot at resonance.  For both 
cases, $S_{11}=0$ in the absence of pumping voltage. Therefore, from 
Eqs.(\ref{see}) we have 
\begin{equation}
\partial_{V_{1/2}} S_{ee} = \partial_{V_{1/2}} 
S_{11} - S_{12}^2 \partial_{V_{2/1}} S_{11}^*
\end{equation}
where we have used the fact that $\partial_{V_1} S_{22} = 
\partial_{V_2} S_{11}$. Using Fisher-Lee relation\cite{lee} 
$S_{\alpha \beta} = -
\delta_{\alpha \beta} + i \sqrt{v_\alpha v_\beta} G^r_{\alpha \beta}$ 
and the Dyson equation $\partial_{V_j} G^r_{\alpha \beta} =
G^r_{\alpha j} G^r_{j \beta}$\cite{gasparian}, we have 
$\partial_{V_1} S_{11} =ivG^r_{11} G^r_{11}=-i/(2k)$ and $\partial_{V_2}
S_{11} = iv G^r_{12} G^r_{21} = -i/(2k) S_{12}^2$ with the velocity
$v=2k$. So for both cases (a) and (b) we have 
constructive interference between direct reflection and the
multiple Andreev reflection $\partial_{V_j} S_{ee} = 2 \partial_{V_j} 
S_{11}$ which gives a pumped current $I_p^{NS}=4 I_p^N$ with
$I_p^N = \pm \omega V_p^2 \sin(4ka)/(4\pi k^2)$, where the plus sign
is for quantum point contact since $S_{12}=e^{2ika}$, and the minus sign
corresponds to resonant tunneling since $S_{12}= e^{-2ika}$. 

In the general situation, the pumped current can be calculated 
numerically using Eq.(\ref{current1})\cite{foot5}. Since the pumped 
current is proportional to $\omega$, we set $\omega=1$ for 
convenience. In the left inset of Fig.1, we plot 
the ratio $I_p^{NS}/I_p^N$ as a function 
of pumping strength $V_p/V_0$ for $V_0=79.2$ at the resonant point. 
We see that as the pumping amplitude increases, the constructive
interference effect is suppressed. At small pumping
strength, the ratio is about four which agrees with our theoretical
analysis. At larger pumping strength, this ratio decreases to the
value below two.
%It shows that $S_{he}$ will not contribute for other $E$. 
%In the inset of Fig.1 we show the pumped currents and
%their ratio for the quantum point contact. 
Similar behaviour is seen for the quantum point contact.
%%%
Fig.1 shows the pumped current as a function of Fermi energy $E_F$
for different pumping amplitudes. For $V_0=79.2$, 
we have chosen $v_g=-9.39$ so that one resonant level in the quantum dot 
is aligned with the chemical potential $\mu_s$ of the superconducting lead 
in the absence of pumping voltage $V_p$.  For comparison, we also plot the 
Andreev reflection coefficient when $V_p=0$. Several observations are 
in order: (1). the pumped current is peaked near the resonant level showing 
clearly a resonant behaviour. This is because the pumped current 
(Eq.(\ref{current1}) is proportional to DOS of the system which reaches 
its maximum near the resonance. (2) As the pumping amplitude $V_p$ increases,
the pumped current increases. (3). The pumped current has two asymmetric 
peaks. To understand this, we plot in the right inset of Fig.1 the Andreev 
reflection coefficient versus $E_F$ at different moments in one pumping 
cycle. From this inset, we observe that the Andreev reflection coefficient 
gives one or two peaks depending on the configuration of the system. This 
behaviour can be understood from the Breit-Wigner form of the resonant 
Anfreev reflection $T_A$ through a single level $E_0=0$ (measured relative 
to $\mu_s$)\cite{wei2}:  
\begin{equation}
T_A=\frac{\Gamma_L^2 \Gamma_R^2} {4(E^2+\Gamma \Delta_\Gamma/4)^2
+\Gamma_L^2 \Gamma_R^2}
\label{bw}
\end{equation}
where $\Delta_\Gamma=\Gamma_L-\Gamma_R$ and $\Gamma=\Gamma_L+\Gamma_R$. 
We see that $T_A$ shows two peaks when $\Gamma_L<\Gamma_R$ and just 
one peak otherwise. Note that in one pumping cycle half of the 
configurations corresponds to $\Gamma_L < \Gamma_R$, therefore two 
pumped current peaks show up in Fig.1 because from Eq.(\ref{current1}) 
the pumped current is obtained through integral over 
all the configurations in one pumping cycle. Finally, the reason that two 
peaks are asymmetric is mainly due to the energy dependence of the
self energy. If $\mu_s$ is right in the middle of two resonant levels 
($E_1$ and $E_2$), {\it i.e.}, $\mu = (E_1+E_2)/2$, then electron coming from 
normal lead with incident energy $E_1$ tunnels into the structure through the 
resonant level $E_1$ and Andreev reflected as a hole back to the quantum 
dot through the resonant level $E_2$ with a Copper pair created in the 
superconducting lead. We now examine the Andreev assisted pumping through 
two levels. For $V_0=79.2$ and $v_g=-23.48$, there are two resonant levels inside 
the subgap ($\mu_s=0$) at $E_1=14.09$ and $E_2=-14.09$.
Hence, strong Andreev reflections can occur near $E_F = E_1$.
Fig.2 shows the two level pumped current versus Fermi energy (solid
line). For comparison we also plot the corresponding Andreev reflection 
coefficient versus Fermi energy when the $V_p$ is switched 
off (see inset of Fig.2). Similar to Fig.1, the pumped current also
shows strong resonant behaviour with smaller amplitude (compare Fig.1
dot-dashed line). We found two peaks of pumped 
current around $E_1$, one is near the resonant energy and the other one 
is shifted to a smaller energy with larger current. Although it is 
similar to Fig.1 but has different origin. This is due to the fact that 
when the pumping gate is turned on, the barrier heights and hence the 
resonant level change with time. To confirm this, we also plot the Andreev 
reflection coefficients $T_A$ at several instants of one pumping
cycle in the same figure. The peaks of $T_A$ shift around the energy level 
at different pumping time and give the behaviour of pumped current. 
We have also calculated the pumped current for other system parameters 
and confirmed that the behaviour of pumped current shown here is generic. 

In summary, in the presence of superconducting lead, the pumped current 
is greatly enhanced due to the quantum interference of direct
reflection and multiple Andreev reflection. In the weak pumping
regime, we have the relation $I_p^{NS} = 4 I_p^N$ for both quantum
point contact and the resonant tunneling structure. Interesting
Andreev assisted pumping behaviours are revealed as well.

We gratefully acknowledge support by a RGC grant from the SAR Government of 
Hong Kong under grant number HKU 7215/99P. H.G. is supported by NSERC of 
Canada and FCAR of Qu\'ebec.

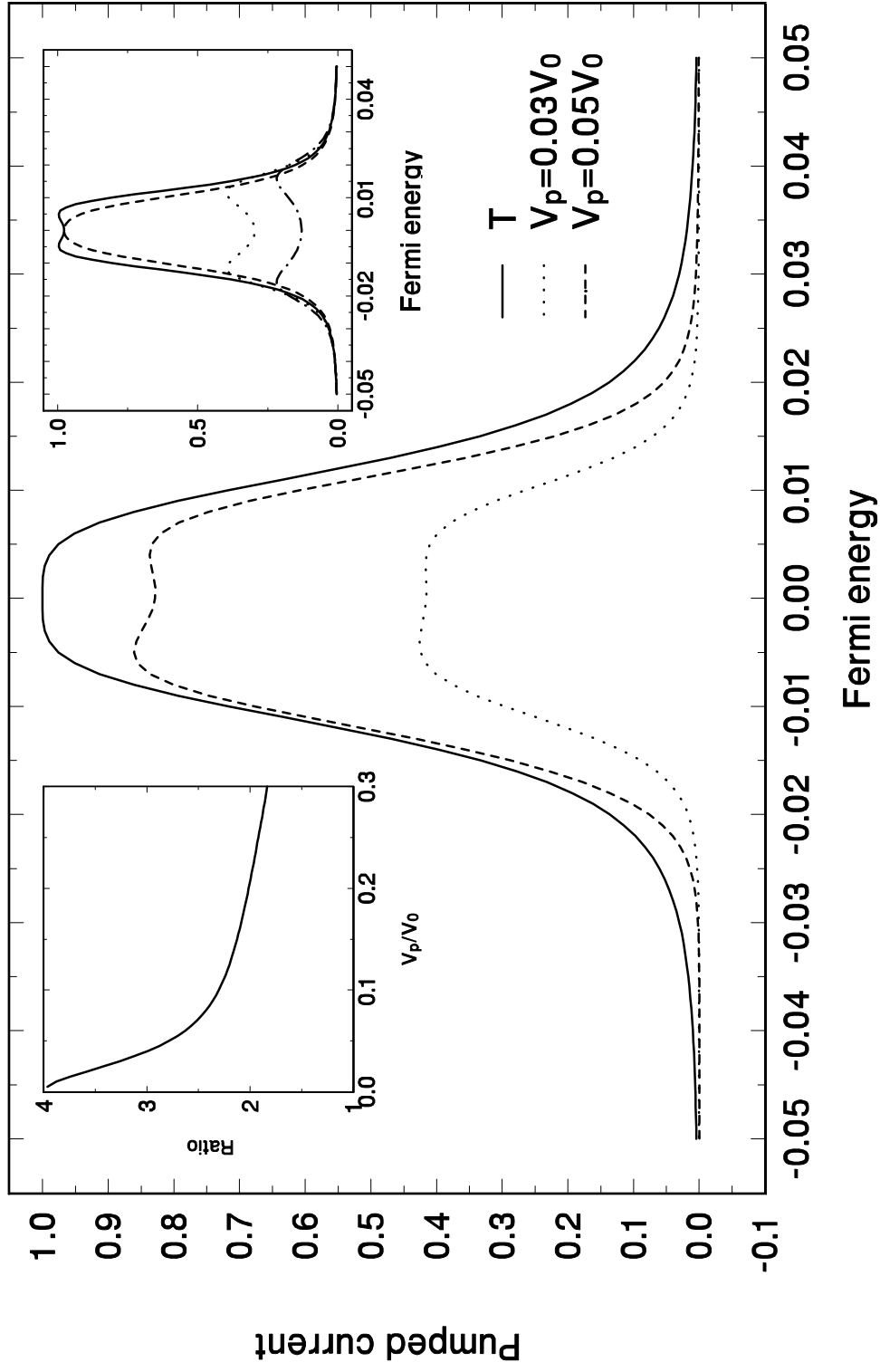
\begin{figure}
\caption{
The pumped current versus $E_F$
at different pumping amplitudes: $V_p=0.03V_0$, $0.05V_0$.
The left inset: $I_p^{NS}/I_p^N$ versus $V_p/V_0$ at resonant point. 
The right inset: Andreev reflection coefficient $T_A$ as a function of $E_F$
with $V_p =0.05V_0$ at different pumping time: $t=\pi/4$ 
(dot dashed line), $t=\pi/2$ (dotted line), $t=3\pi/4$ (solid line), and 
$t=7\pi/4$ (dashed line). System parameters in Fig.1 and the insets: 
$V_0=79.2$, $v_g=-9.39$, $\phi=\pi/2$, and $\Delta=20$.
}
\end{figure}

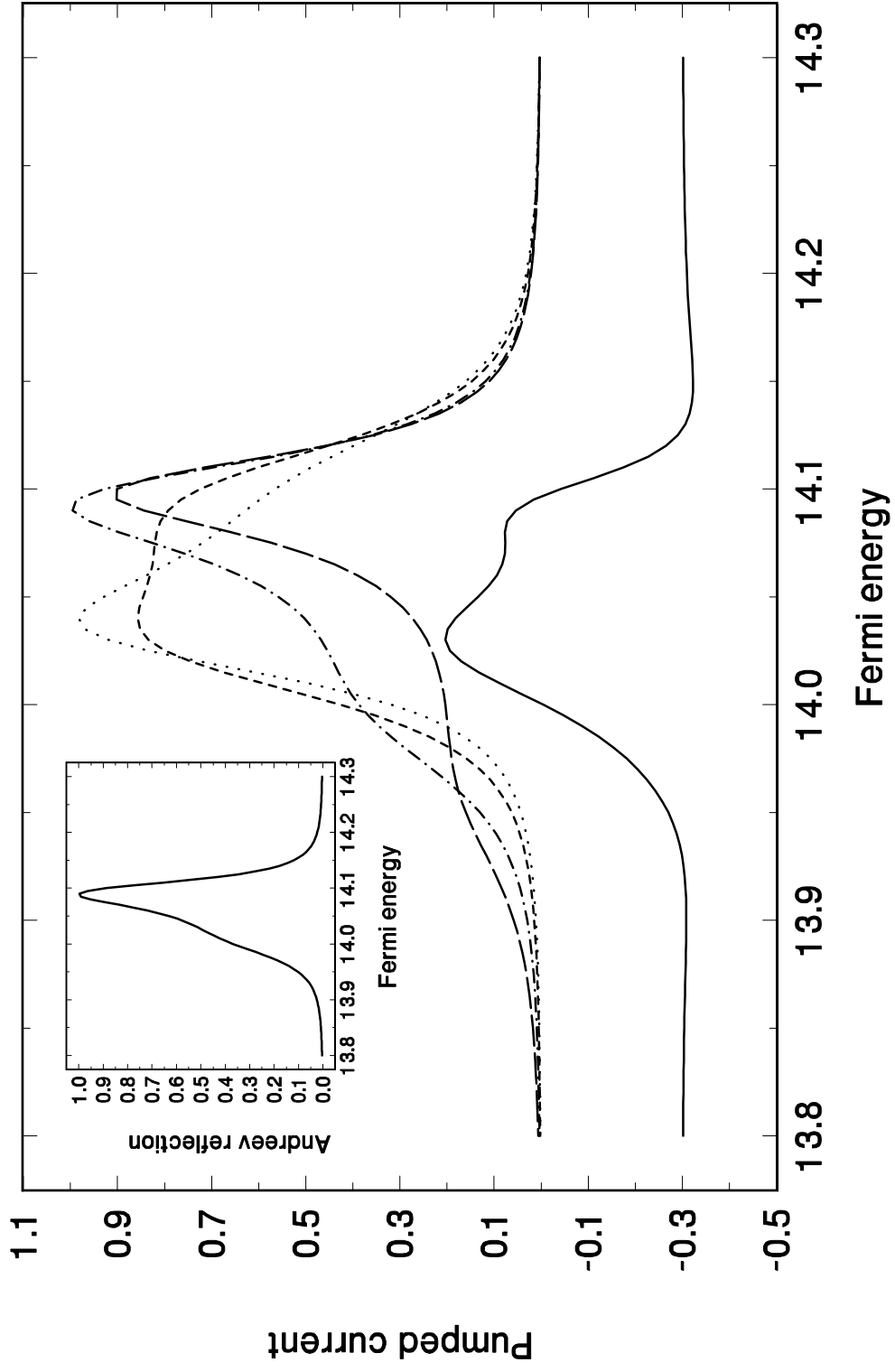
\begin{figure}
\caption{
The two level pumped current $I_p$ as a function of Fermi energy (solid 
line) with $V_p=0.05V_0$ and $v_g=-23.481$. We also plot $T_A$ at several 
pumping time: $t=\pi/4$ (dotted line), $t=\pi/2$ (dashed line), $t=3\pi/4$ 
(dot-dashed line), and $t=\pi$ (long dashed line). The pumped
current has been offset by 0.3 for illustrating purpose. Inset: $T_A$ 
versus Fermi energy when $V_p=0$.  Other parameters are the same as Fig.1.
}
\end{figure}


\begin{thebibliography}{00}
\bibitem{foot4}
For the GaAs system with $a=1000A$, the energy uint is $E=56 \mu eV$.
\bibitem{foot1}
This formula can be derived using the Keldysh nonequilibrium Green's 
function method used in the present work.
\bibitem{foot2}
A factor of two has been included for spin.
\bibitem{foot3}
Here we assume a zero magnetic field. Since $S_{12}$ is symmetric with 
respect to $V_1$ and $V_2$, it does not contribute to $I_p^N$.
\end{thebibliography}
\end{document}